\documentclass[journal,10pt,twocolumn]{IEEEtran}
\usepackage{graphics}
\usepackage{epsfig}
\usepackage{amsmath}
\usepackage{cite}
\usepackage{setspace}
\usepackage{algorithm}
\usepackage{algorithmic}
\usepackage{bm}
\usepackage{color}

\usepackage{amssymb}  
\usepackage{booktabs}
\usepackage{tabularx}
\usepackage{makecell}

\IEEEoverridecommandlockouts                              
\title{Secure Fusion Estimation Against FDI Sensor Attacks in Cyber-Physical Systems}
\author{Bo Chen,~Pindi Weng,~Daniel W.C. Ho~and Li Yu
\thanks{B. Chen, P. Weng and L. Yu are with the Department of Automation, Zhejiang University of Technology, Hangzhou 310023, China (email: bchen@aliyun.com).}
\thanks{D. W. C. Ho is with the Department of Mathematics, City University of
Hong Kong, Hong Kong, 999077.}
}

\begin{document}

\markboth{}
{Shell \MakeLowercase{\textit{et al.}}: Bare Demo of IEEEtran.cls for Journals}
\maketitle

\begin{abstract}
This paper is concerned with the problem of secure multi-sensors fusion estimation for cyber-physical systems, where sensor measurements may be tampered with by false data injection (FDI) attacks. In this work, it is considered that the adversary may not be able to attack all sensors. That is, several sensors remain not being attacked. In this case, new local reorganized subsystems including the FDI attack signals and un-attacked sensor measurements are constructed by the augmentation method. Then, a joint Kalman fusion estimator is designed under linear minimum variance sense to estimate the system state and FDI attack signals simultaneously. Finally, illustrative examples are employed to show the effectiveness and advantages of the proposed methods.
\end{abstract}

\section{Introduction}
Cyber-physical systems (CPSs) are intellectualized complex systems that combine the computing, the network communications and the physical environment. With the help of communication networks, key facilities are integrated by CPSs, which makes the interaction between the cyberspace and the physical world more convenient {\cite{c10,c1,c18,c100}}. Therefore, CPSs have attracted wide attentions and have been applied in various fields such as the intelligent transportation, the smart grids, the medical and healthcare systems and the process automation systems \cite{c999,c2,c3}. As a key issue in CPSs, the real-time state estimation based on sensor measurements plays a crucial role for providing CPSs with the real-time monitoring and control capability \cite{c888}. Take the power system as an example, the state estimation results can be utilized for fulfilling power system control and real-time contingency analysis \cite{c777}. In this case, the accuracy of state estimation has an important impact on the safe and efficient operation of CPSs \cite{c666}. For this reason, the multi-sensors fusion estimation, which can potentially improve estimation accuracy and enhance robustness, has been studied in \cite{c998,c997,c96,c97,c98} for different CPSs.

Generally, the closure of the system is broken in CPSs due to the opening of communication networks. This makes the system face threats from cyber-attacks \cite{c11}, such as the denial-of-service (DoS) attacks and the false data injection (FDI) attacks \cite{c996}. Particularly, the FDI attacks are able to tamper the measurement signals transmitted by the communication networks. Then, traditional measurement-based state estimation methods cannot perform well based on the tampered measurements, which degrades the estimation performance for CPSs. As a result, successful FDI attacks may cause serious industrial accidents and economic losses \cite{c13}. Therefore, the secure state estimation which estimates the system state from compromised measurements has become one of the vital research directions \cite{c57,c62,c54,c61}. Also, secure estimation problem was solved in \cite{c62} by formulating it into a classical error correction problem, and the secure state estimation method was combined with Kalman filter to improve the estimation performance. In \cite{c61}, prior information was utilized to reinforce the system resilience against malicious sensor attacks, and then an intermediate-variable-based estimation method was developed in \cite{c995} to estimate FDI attacks occurring at the actuator and the sensor in CPSs. Notice that the aforementioned methods only consider the single-sensor condition, however, multi-sensor fusion can provide more redundant information for guaranteeing the security and accuracy of estimation algorithms.

Under the case of multi-sensor, secure state estimation methods can be divided into two categories. The first class of methods is to detect the attack signals and then weaken the impact caused by the attacks. For instance, a finite-time horizon detector was proposed in \cite{c58} to solve the attack detection problem, then an event-driven supervised estimator was designed to guarantee the security of estimation performance. In \cite{c53}, a distributed adaptive algorithm based on Kullback-Leibler divergence was proposed to detect FDI attacks, and then three different algorithms were explored separately to weaken the impact of attacks. Meanwhile, the secure state estimation problem was solved in \cite{c60} by a trust-based diffusion algorithm with adaptive combination policy. Then, a Gaussian-mixture-model-based detection algorithm was developed in \cite{c52} which can fuse measurements from different sensors accordingly based on a belief provided for each sensor. It should be pointed out that the detection accuracy of FDI attack signals in those works is dependent on the detection threshold, but how to determine the most reasonable detection threshold is always a difficult problem.

Different from the processing idea in the first class of methods, the second class of methods is to directly estimate the system state and the FDI attack signal simultaneously, which can avoid the design of detection threshold. In \cite{c55}, a projected sliding-mode observer-based estimation algorithm was developed to reconstruct the system state from the sensor measurements corrupted by malicious attacks. Subsequently, a novel secure Luenberger-like observer was designed in \cite{c54} to estimate the state and attacks from the tampered measurements. In \cite{c56}, a switched Luenberger observer with a projection operator was proposed to estimate the state of an augmented system, where the augmented system was constructed by treating attacks as parts of the state. Meanwhile, a switched gradient descent technique was used in \cite{c63} to develop a novel algorithm that can deal with the secure state estimation problem. In \cite{c59}, the attacked CPS was modeled as a finite-state hidden Markov model with switching transition probability matrices, based on which a joint state and attack estimation method was proposed. Notice that, from the perspective of information fusion, the above-mentioned methods were studied under the framework of centralized fusion, i.e., measurements from different sensors were modeled as a high-dimensional measurement. In fact, the centralized fusion structure has poor robustness and reliability when there 
{is a faulty fusion center}, while the distributed fusion structure is generally more robust, reliable, and fault tolerant \cite{c96,c99}. However, few results focus on the second class of methods under distributed fusion framework.
\begin{figure}
\centering
\includegraphics[width=0.40\textwidth]{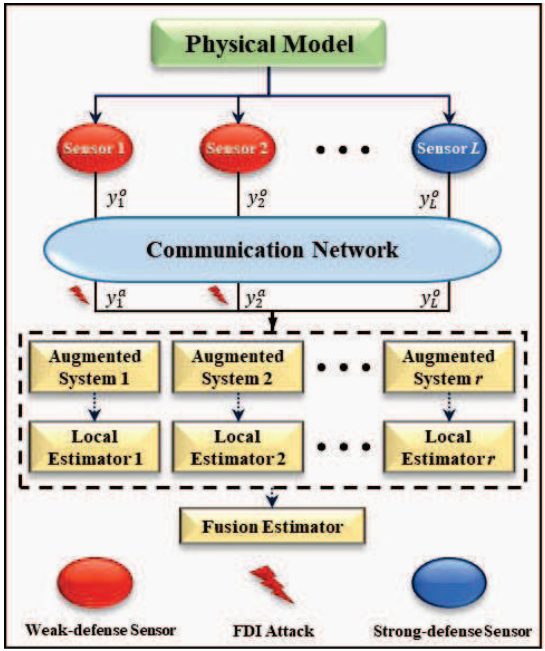}
{\caption{When each sensor node sends measurement information to the monitoring center, FDI attacks may exist during the transmission. To estimate states of the system, augmented systems are constructed and local/fusion estimators are designed based on the augmented systems.}}
\label{fig1}
\end{figure}

Motivated by the aforementioned analysis, this paper shall study the secure fusion estimation methods to simultaneously estimate the CPSs' states and the FDI attack signals under the distributed fusion framework. In this paper, it is considered that sensor measurements may be corrupted by FDI attacks. It should be pointed out that the number of attacked sensors is not limited and the prior information of attacks is not required to be known. The main contributions of this paper can be summarized as follows:
\begin{itemize}
\item  A new reorganized subsystem model based on un-attacked sensor measurements is constructed by augmenting the FDI attack signals into the system state vector, where the difference of the attacks between the current moment and the previous moment is modeled as an unknown input in this new model. Based on the constructed model, an efficiently joint local estimation structure is proposed to simultaneously estimate the new system states and unknown inputs. Then, a uniform structure of distributed fusion estimators is proposed to fuse the local information generated from the local estimators of CPSs' states and FDI attack signals.
\item Optimal local joint estimators, which can simultaneously estimate the system states and the FDI attack signals, are designed in the linear minimum variance sense. In this method, the compensation factor is proposed to adjust the estimation performance by compensating the unknown term with respect to attack signals. According to the designed local joint estimator, distributed fusion criteria based on the multi-sensor information are designed by using the matrix-weighted fusion methods.
\end{itemize}
Finally, illustrative examples are employed to show the advantages and effectiveness of the proposed methods.

\indent Notations: $\mathbb{R}^r$ and $\mathbb{R}^{r\times s}$ denote the $r$-dimensional and $r\times s$ dimensional Euclidean spaces, respectively. $\mathrm{E}\{\cdot\}$ denotes mathematical expectation, while $\mathrm{diag}\{\cdot\}$ stands for a block diagonal matrix. `$I$' represents the identity matrix with appropriate dimensions and `$O$' is zero matrix. The superscript `$\mathrm{T}$' represents the transpose, while $X>(<)\ 0$ denotes a positive-definite (negative-definite) matrix. $\mathrm{Tr}(\cdot)$ represents the trace of the matrix.

\section{Problem Formulations}
Consider a physical process monitored by $L$ sensors (see Fig. 1), where the physical process and sensor measurements are modeled by:
\begin{equation}
\label{eq:1}
\begin{cases}
\bm{x}(k)=A(k)\bm{x}(k-1)+\bm{w}(k-1)\\
\bm{y}^o_i(k)=C_i^o(k)\bm{x}(k)+\bm{v}^o_i(k), i=1,2,\ldots,L
\end{cases}
\end{equation}
where $\bm{x}(k)\in\mathbb{R}^n$ is the system state, $\bm{y}^o_i(k)\in\mathbb{R}^{p_i}$ is the measurement of the $i$th sensor. $A(k)$ and $C^o_i(k)$ are known matrices. $\bm{w}(k)$ and $\bm{v}^o_i(k)$ are zero-mean Gaussian white noises with known covariance $Q$ and $R^o_i$.

When sensor measurements are transmitted to the monitoring center over communication networks, an adversary is able to launch FDI attacks to tamper measurement signals. However, it is not practical and not economical for the adversary to attack all sensors. In this sense, it is considered in this paper that the adversary may attack several sensors while the other sensors are completely secure.

\textbf{Definition 1.} (Strong/weak-defense sensor) The sensors that may be attacked by the adversary are defined as weak-defense sensors, and the sensors that are well protected from FDI attacks are defined as strong-defense sensors.

According to Definition 1, it is specified that the first $r$ sensors are arranged as the weak-defense sensors, while the last $L-r$ are strong-defense sensors, i.e., the measurement $\bm{y}_i^o(k)\ (i=r+1,\ldots,L)$ will not be tampered.

Let the $i$th attacked measurement be $\bm{y}_i^a(k)$, then $\bm{y}_i^a(k)$ is modeled by:
\begin{equation}
\label{eq:3}
\bm{y}_i^a(k)=\bm{y}^o_i(k)+\bm{\theta}_i(k),\ i=1,2,\ldots,r
\end{equation}
where $\bm{\theta}_i(k)\in\mathbb{R}^{p_i}$ is the FDI attack signal. To estimate the system state and FDI attack signals accurately, the weak-defense sensors are combined with strong-defense sensors, which leads to
\begin{equation}
\label{eq:80}
\begin{cases}
C_i(k)\triangleq[C^o_i(k);C^o_{j}(k);\ldots;C^o_{j_o}(k)]\\
\bm{v}_i(k)\triangleq[\bm{v}^o_i(k);\bm{v}^o_{j}(k);\ldots;\bm{v}^o_{j_o}(k)]
\end{cases}
\end{equation}
where {$j,\ldots,j_o\in\{r+1,\ldots,L\}$}, and it yields the enhanced measurement as follows
\begin{equation}
\label{eq:66}
\bm{y}_i(k)=C_i(k)\bm{x}(k)+\Phi_i\bm{\theta}_i(k)+\bm{v}_i(k)\in\mathbb{R}^{m_i}
\end{equation}
where $\Phi_i\triangleq[I_{p_i};O_{p_{j}\times p_i};\ldots;O_{p_{j_o}\times p_i}]$, this indicates that the weak-defense sensor $i$ may be attacked, while the strong-defense sensors are secure. Subsequently, define $\bm{X}_i(k)\triangleq[\bm{x}(k);\bm{\theta}_i(k)]$, and a new augmented system is given by:
\begin{equation}
\label{eq:4}
\begin{cases}
\bm{X}_i(k)=A^a_i(k)\bm{X}_i(k-1)+\Phi^a_i\bm{\phi}_i(k)\\
\ \ \ \ \ \ \ \ \ \ \ \ \ +\bm{W}_i(k-1) \\
\bm{y}_i(k)=C^a_i(k)\bm{X}_i(k)+\bm{v}_i(k)
\end{cases}
\end{equation}
where $i=1,2,\ldots,r$ and
\begin{equation*}
\begin{cases}
A^a_i(k)\triangleq{\rm diag} \{ A(k),I_{p_i}\}
\\\bm{\phi}_i(k)\triangleq\bm{\theta}_i(k)-\bm{\theta}_i(k-1)\in\mathbb{ R}^{p_i}
\\\Phi^a_i\triangleq[O_{n\times p_i};I_{p_i}]
\\\bm{W}_i(k-1)\triangleq[\bm{w}(k-1);O_{p_i\times1}]
\\C_i^a(k)\triangleq[C_i(k),\Phi_i]
\end{cases}\;\;\;\;\;\;
\end{equation*}
The augmented system state shall be observable based on the sensor meausurement at each time to obtain satisfactory estimation performance.

Based on the measurements $\{\bm{y}_i(1),\ldots,\bm{y}_i(k)\}$, it is proposed in this paper that the state $\bm{X}_i(k)$ including attack signals and the input signal $\bm{\phi}_i(k)$ can be estimated jointly by the following recursive form \cite{c95}
\begin{equation}
\label{eq:99}
\begin{cases}
\hat{\bm{X}}_i(k)=A^a_i(k)\hat{\bm{X}}_i(k-1)+\Phi^a_i\hat{\bm{\phi}}_i(k-1)
\\\ \ \ \ \ \ \ \ \ \ \ \ \ +K_i(k)\tilde{\bm{y}}_i(k)
\\\hat{\bm{\phi}}_i(k)=\hat{\bm{\phi}}_i(k-1)+\Gamma_i(k)\tilde{\bm{y}}_i(k)
\end{cases}
\end{equation}
where
\begin{equation}
\label{eq:2}
\begin{aligned}
&\tilde{\bm{y}}_i(k)\triangleq\bm{y}_i(k)-C^a_i(k)[A^a_i(k)\hat{\bm{X}}_i(k-1)
\\&\ \ \ \ \ \ \ \ \ \ \ +\Phi^a_i\hat{\bm{\phi}}_i(k-1)]
\end{aligned}
\end{equation}
Here, $\hat{ \bm{X}}_i(k)$ and $\hat{ \bm{\phi}}_i(k)$ are local estimates, while $K_i(k)$ and $\Gamma_i(k)$ are the gains to be designed. Under the framework of distributed fusion, the fusion state estimator is given by:
\begin{equation}
\label{eq:6}
\hat{\bm{x}}_0(k)=\sum^r_{i=1}G_i(k)\hat{\bm{x}}_i(k)
\end{equation}
where $\hat{\bm{x}}_i(k)\triangleq[I_n, O_{n\times p_i}]\hat{\bm{X}}_i(k)$, and each $G_i(k)$ is the weight to be designed, which satisfies $\sum^r_{i=1}G_i(k)=I_n$.\\
\indent
Consequently, the aim of this paper is to design optimal gains $K_i(k)$, $\Gamma_i(k)$ in (\ref{eq:99}) and each weighting fusion matrix $G_i(k)$ in (\ref{eq:6}) in linear minimum variance sense.

\textbf{Remark 1.} Under the centralized framework, $s$-sparse attacks of sensor measurement $y \in \mathbb{R}^m$ were considered in [20] and [27]-[29], where the number $s$ of the attacked elements were required to satisfy $s\le (m/2-1)$, and then the system states can still be estimated from the tampered sensor measurement. In this sense, $m$ may be a large value for multi-sensor fusion systems, which means that a large number of sensors are supposed not being attacked. Different from the above-mentioned attack schemes, under the distributed fusion framework, the strong-defense sensors are proposed in this paper to play helpful roles in assisting the weak-defense sensor. Then, only a few sensors are required to be protected well from attacks, and thus the defense cost can be reduced. {On the other hand, the augmentation method in this paper is not efficient under the framework of centralized fusion, because the dimension of the system state increases when the number of sensors is large. This brings a huge amount of computation. However, for the distributed fusion in this paper where the augmented system (\ref{eq:4}) is constructed for each sensor measurement, the state of each augmented system $i$ only contains the original system state and the attack signal of sensor $i$. Thus, the computation for each augmented system with low dimension is not huge, despite a large number of sensors.}

{\textbf{Remark 2.} Existing attack detection methods in \cite{c58,c53,c60,c52} can be utilized to confirm which sensors are not under attack. In this case, by implementing a specific attack detection method, the $L-r$ sensors with the highest confidence level are viewed as the strong-defense sensors (i.e. the sensors that are not tampered with by FDI attacks). Note that, for the first class of methods, the detection threshold should be chosen ``properly'', otherwise the attacked sensor cannot be detected (the threshold is too large) or false alarm arises (the threshold is too small). However, in this paper, the detection methods are merely utilized to confirm the strong-defense sensors. Thus, the threshold can be a small value such that only the sensors with a high confidence level are viewed as not being attacked.}

\textbf{Remark 3.} For the augmented system (\ref{eq:4}), a direct way is treating the term $\Phi^a_i\bm{\phi}_i(k)$ as the noise. Then Kalman filter can be used to estimate the augmented state $\bm{X}_i(k)$
\begin{equation}
\label{eq:98}
\begin{aligned}
&\hat{\bm{X}}_i(k)=A^a_i(k)\hat{\bm{X}}_i(k-1)+K_i^f(k)[\bm{y}_i(k)\\
&\ \ \ \ \  \ \ \ \ \  -C^a_i(k)A^a_i(k)\hat{\bm{X}}_i(k-1)]
\end{aligned}
\end{equation}
where $K_i^f(k)$ is the gain matrix obtained by Kalman filter. However, since there is no statistical information about the signal $\bm{\phi}_i(k)$, the standard Kalman filter cannot work well. Moreover, the advantages of the proposed methods in this paper have been demonstrated by comparing with the above direct method in Simulations.

\section{Main Results}
Before deriving the main results, define:
\begin{equation}
\label{eq:n5}
\begin{cases}
Q^a_i\triangleq{\rm diag}\{Q,O_{p_i\times p_i}\}
\\R_i\triangleq{\rm diag}\{R^ o _i ,R^ o _{r+1} ,...,R^ o _{L}\}
\\Q^a_{ij}\triangleq[Q,O_{n\times p_j};O_{p_i\times n},O_{p_i\times p_j}]
\\\Gamma^a_i(k)\triangleq I_{p_i}-\Gamma_i(k)C^a_i(k)\Phi^a_i
\\\Gamma^b_i(k)\triangleq\Gamma_i(k)C^a_i(k)A^a_i(k)
\\K^a_i(k)\triangleq I_{n+p_i}-K_i(k)C^a_i(k)
\end{cases}
\end{equation}
and
\begin{equation}
\label{eq:5}
\begin{cases}
\tilde{\bm{\phi}}_i(k)\triangleq\bm{\phi}_i(k)-\hat{\bm{\phi}}_i(k)
\\\tilde{\bm{X}}_i(k)\triangleq\bm{X}_i(k)-\hat{\bm{X}}_i(k)
\\P^\phi_{ij}(k)\triangleq{\rm E}\{\tilde{\bm{\phi}}_i(k)\tilde{\bm{\phi}}^{\rm T}_j(k)\}
\\P^X_{ij}(k)\triangleq{\rm E}\{\tilde{\bm{X}}_i(k)\tilde{\bm{X}}^{\rm T}_j(k)\}
\\\Psi_{ij}(k)\triangleq{\rm E}\{\tilde{\bm{X}}_i(k)\tilde{\bm{\phi}}^{\rm T}_j(k)\}
\\U_{ij}(k)\triangleq{\rm E}\{\tilde{\bm{X}}_i(k)\hat{\bm{\phi}}^{\rm T}_j(k)\}
\\Y_{ij}(k)\triangleq{\rm E}\{\tilde{\bm{\phi}}_i(k)\hat{\bm{\phi}}^{\rm T}_j(k)\}
\\V_{ij}(k)\triangleq{\rm E}\{\hat{\bm{\phi}}_i(k)\hat{\bm{\phi}}^{\rm T}_j(k)\}
\\P^\theta_{ij}(k)\triangleq{\rm E}\{{\bm{\theta}}_i(k){\bm{\theta}}^{\rm T}_j(k)\}
\end{cases}\;\;\;\;\;\;\;\;
\end{equation}

According to the results in \cite{c99}, a group of optimal weighting matrices $G_i(k)\ (i=1,\ldots,r)$ in (\ref{eq:6}) can be determined in the linear minimum variance sense by the following form:
\begin{equation}
\label{eq:21}
G(k)=\Sigma^{-1}(k)H(H^{\rm T}\Sigma^{-1}(k)H)^{-1}
\end{equation}
where
\begin{equation}
\label{eq:22}
\begin{cases}
G(k)\triangleq[G^{\rm T}_1(k);\ldots;G_r^{\rm T}(k)]\in\mathbb{ R}^{nr\times n}
\\H\triangleq[I_{n};\ldots;I_{n}]\in\mathbb{ R}^{nr\times n}
\\\Sigma(k)\triangleq\{P^x_{ij}(k)\}\in\mathbb{ R}^{nr\times nr}\\
P^x_{ij}(k)\triangleq[I_n,O_{n\times p_i}]P^X_{ij}(k)[I_n;O_{p_j\times n}]
\end{cases}
\end{equation}
It follows from (\ref{eq:21}) and (\ref{eq:22}) that covariance matrices $P^X_{ij}(k)\ (\forall i,j)$ are needed, while $P_{ij}^X(k)$ is determined by $\Gamma_i(k)$ and $K_i(k)$. In this case, the estimator gains $\Gamma_i(k)$, $K_i(k)$ and the local estimation error covariance will be given by Theorem 1 and Lemma 1, while the estimation error cross-covariance will be presented by Theorem 2.

{Notice that $\bm{\theta}_i(k)$ is the attack signal generated from the adversary and no assumption is made on it in this paper. In this case, $\bm{\theta}_i(k)$ can be a random signal or it may not obey a probabilistic law, which is designed by the attacker and is unknown to the defender. In this subsection, $\bm{\theta}_i(k)$ is treated as a random signal to calculate the covariance matrices. However, since it is difficult for the defender to obtain the correlation of each attack signal with the previous system states, the previous attacks and the attack injected into another sensor, the following general situation is considered:}
\begin{eqnarray}
\label{eq:666}
\begin{cases}
{\rm E}\{\bm{\theta}_i(k){\bm{\theta}}^{\rm T}_j(k)\}=O_{p_i\times p_j}(i\ne j)
\\{\rm E}\{\bm{\theta}_i(k){\bm{X}}^{\rm T}_j(t)\}=O_{p_i\times (n+p_j)}
\\{\rm E}\{\bm{\theta}_i(k)\hat{\bm{\phi}}^{\rm T}_j(t)\}=O_{p_i\times p_j}
\\{\rm E}\{\bm{\theta}_i(k)\hat{\bm{X}}^{\rm T}_j(t)\}=O_{p_i\times (n+p_j)}
\end{cases}
\end{eqnarray}
where $t=0,\cdots,k-1$.

{\textbf{Remark 4.} Notice that the attack signals designed by the adversary may satisfy a certain rule and the defender can estimate the attacks well if the rule is available. In fact, it is difficult for the defenders to know the attack information, and the right sides of equations in (14) shall be unknown matrices depending on $k,t,i,j$. In this paper, the condition (\ref{eq:666}) is considered and it can be seen as the worst case that the influence of correlations to the calculation of the covariance matrices is ignored. To improve the estimation performance, the compensation factor will be proposed later, which can potentially compensate the unknown covariance information on the attacks.}

Under the condition (\ref{eq:666}), the recursive form of each local estimation error covariance is first presented in Lemma 1.

\indent \textbf{Lemma 1.} Under the initial values $P^\phi_{ii}(0)$, $P^X_{ii}(0)$, $U_{ii}(0)$ and $V_{ii}(0)$. Suppose that the compensation factor $\eta_i\ge0$ and estimator gains $K_i(k)$, $\Gamma_i(k)$ are given, then the matrices $P^\phi_{ii}(k)$, $P^X_{ii}(k)$, $U_{ii}(k)$ and $V_{ii}(k)$ can be calculated by:
\begin{eqnarray}
\label{eq:10}
\begin{array}{l}
\;P_{ii}^\phi (k) = \Gamma _i^a(k)\Xi _i^1(k) - \Xi _i^1(k){\{ {\Gamma _i}(k)C_i^a(k)\Phi _i^a\} ^{\rm{T}}}\\
\;\;\;\;\; + {\{ \Gamma _i^b(k)\Xi _i^2(k)\} ^{\rm{T}}} + \Gamma _i^b(k)\Xi _i^2(k) + {\Gamma _i}(k){R_i}\Gamma _i^{\rm{T}}(k)\;\\
\;\;\;\;\; + {\Gamma _i}(k)C_i^a(k){\Xi _i}(k){\{ {\Gamma _i}(k)C_i^a(k)\} ^{\rm{T}}}
\end{array}
\end{eqnarray}
\begin{eqnarray}
\label{eq:11}
P_{ii}^X(k) = K_i^a(k){\Xi _i}(k){\{ K_i^a(k)\} ^{\rm{T}}} + {K_i}(k){R_i}K_i^{\rm{T}}(k)
\end{eqnarray}
\begin{eqnarray}
\label{eq:12}
\begin{array}{l}
{U_{ii}}(k) = K_i^a(k)[A_i^a(k){U_{ii}}(k - 1) - \Phi _i^a{V_{ii}}(k - 1)]\\
\;\;\;\;\; - {\eta _i}K_i^a(k)\Phi _i^a{\{ {\Gamma _i}(k - 1)C_i^a(k - 1)\Phi _i^a\} ^{\rm{T}}}\\
\;\;\;\;\; - {K_i}(k){R_i}\Gamma _i^{\rm{T}}(k) + K_i^a(k){\Xi _i}(k){\{ {\Gamma _i}(k)C_i^a(k)\} ^{\rm{T}}}
\end{array}
\end{eqnarray}
\begin{eqnarray}
\label{eq:103}
\begin{array}{l}
\;{V_{ii}}(k) = {\{ \Gamma _i^b(k){U_{ii}}(k - 1)\} ^{\rm{T}}} + \Gamma _i^b(k){U_{ii}}(k - 1)\\
\;\;\;\;\; + {V_{ii}}(k - 1){\{ \Gamma _i^a(k)\} ^{\rm{T}}} - {\Gamma _i}(k)C_i^a(k)\Phi _i^a{V_{ii}}(k - 1)\\
\;\;\;\;\; - {\eta _i}{\Gamma _i}(k - 1)C_i^a(k - 1)\Phi _i^a{\{ {\Gamma _i}(k)C_i^a(k)\Phi _i^a\} ^{\rm{T}}}\\
\;\;\;\;\; - {\eta _i}{\Gamma _i}(k)C_i^a(k)\Phi _i^a{\{ {\Gamma _i}(k - 1)C_i^a(k - 1)\Phi _i^a\} ^{\rm{T}}}\\
\;\;\;\;\; + {\Gamma _i}(k)[C_i^a(k){\Xi _i}(k){\{ C_i^a(k)\} ^{\rm{T}}}+R_i]\Gamma _i^{\rm{T}}(k)
\end{array}
\end{eqnarray}
where
\begin{eqnarray}
\label{eq:9}
\begin{cases}
\begin{aligned}
&\Xi_i(k)\triangleq A^a_i(k)P^X_{ii}(k-1)\{A^a_i(k)\}^{\rm T}+Q^a_i\\
&\ \ \ \ \ \ \ \ \ \  +\Phi^a_i\Xi^1_i(k)\{\Phi^a_i\}^{\rm T}-A^a_i(k)\Xi^2_i(k)\{\Phi^a_i\}^{\rm T}\\
&\ \ \ \ \ \ \ \ \ \  -\Phi^a_i\{A^a_i(k)\Xi^2_i(k)\}^{\rm T}\\
&\Xi^1_i(k)\triangleq 6\eta_iI_{p_i}-P^\phi_{ii}(k-1)-\eta_i\{\Gamma^a_i(k-1)\}^{\rm T}\\
&\ \ \ \ \ \ \ \ \ \  -\eta_i{\Gamma^a_i}(k-1)\\
&\Xi^2_i(k)\triangleq U_{ii}(k-1)+\eta_i{K^a_i}(k-1){\Phi^a_i}
\end{aligned}
\end{cases}
\end{eqnarray}
and $\Gamma^a_i(k),\Gamma^b_i(k), K^a_i(k),Q_i^a,R_i$ are defined in (\ref{eq:n5}).

\textbf{Proof.} Define $\bm{\mu}_i(k-1)\triangleq\bm{\phi}_i(k)-\bm{\phi}_i(k-1)$. Then, the estimation error $\tilde{\bm{\phi}}_i(k)$ defined in (\ref{eq:5}) is given by:
\begin{eqnarray}
\label{eq:888}
\begin{aligned}
&\tilde{\bm{\phi}}_i(k)=[\bm{\phi}_i(k)-\bm{\phi}_i(k-1)]+\bm{\phi}_i(k-1)-\hat{\bm{\phi}}_i(k)\\
&\ \  \ \ \ \ \ =\bm{\mu}_i(k-1)+\tilde{\bm{\phi}}_i(k-1)-\Gamma_i(k)\tilde{\bm{y}}_i(k)
\end{aligned}
\end{eqnarray}
Substituting $\tilde{\bm{y}}_i(k)$ (\ref{eq:2}) into (\ref{eq:888}) yields that
\begin{eqnarray}
\label{eq:13}
\begin{aligned}
&\tilde{\bm{\phi}}_i(k)
 =\Gamma^a_i(k)[\bm{\mu}_i(k-1)+\tilde{\bm{\phi}}_i(k-1)]\\
&\ \ \ \ \ \ \ \ \ \ -\Gamma^b_i(k)\tilde{\bm{X}}_i(k-1)-\Gamma_i(k)\bm{v}_i(k)\\
&\ \ \ \ \ \ \ \ \ \ -\Gamma_i(k)C^a_i(k)\bm{W}_i(k-1)
\end{aligned}
\end{eqnarray}
In the meantime, the estimation error $\tilde{\bm{X}}_i(k)$ in (\ref{eq:5}) can be calculated by:
\begin{eqnarray}
\label{eq:16}
\begin{aligned}
&\tilde{\bm{X}}_i(k)=
 K_i^a(k)[A^a_i(k)\tilde{\bm{X}}_i(k-1)\\
&\ \ \ \ \ \ \ \ \ \ \  +\Phi^a_i\tilde{\bm{\phi}}_i(k-1)+\Phi^a_i\bm{\mu}_i(k-1)\\
&\ \ \ \ \ \ \ \ \ \ \  +\bm{W}_i(k-1)]-K_i(k)\bm{v}_i(k)
\end{aligned}
\end{eqnarray}
where $\Gamma^a_i(k),\Gamma^b_i(k)$ and $K^a_i(k)$ are defined in (\ref{eq:n5}). Then, according to (\ref{eq:5}) and (\ref{eq:13}), the local estimation error covariance matrix $P^\phi_{ii}(k)$ is obtained by
\begin{eqnarray}
\label{eq:14}
\begin{aligned}
&P^\phi_{ii}(k)=\Gamma_i(k)[C^a_i(k)Q^a_i\{C^a_i(k)\}^{\rm T}+R_i]\Gamma_i^{\rm T}(k)\\
&\ \ \ \ \ \ +\Gamma^a_i(k)P^\phi_{ii}(k-1)\{\Gamma^a_i(k)\}^{\rm T}\\
&\ \ \ \ \ \ +\Gamma^b_i(k)P^X_{ii}(k-1)\{\Gamma^b_i(k)\}^{\rm T}\\
&\ \ \ \ \ \ -\Gamma^a_i(k)\Psi^{\rm T}_{ii}(k-1)\{\Gamma^b_i(k)\}^{\rm T}\\
&\ \ \ \ \ \ -\Gamma^b_i(k)\Psi_{ii}(k-1)\{\Gamma^a_i(k)\}^{\rm T}\\
&\ \ \ \ \ \ +\Gamma^a_i(k){\rm E}\{\bm{\mu}_i(k-1)\bm{\mu}^{\rm T}_i(k-1)\}\{\Gamma^a_i(k)\}^{\rm T}\\
&\ \ \ \ \ \ +\Gamma^a_i(k){\rm E}\{\bm{\mu}_i(k-1)\tilde{\bm{\phi}}^{\rm T}_i(k-1)\}\{\Gamma^a_i(k)\}^{\rm T}\\
&\ \ \ \ \ \ -\Gamma^a_i(k){\rm E}\{\bm{\mu}_i(k-1)\tilde{\bm{X}}^{\rm T}_i(k-1)\}\{\Gamma^b_i(k)\}^{\rm T}\\
&\ \ \ \ \ \ +\Gamma^a_i(k){\rm E}\{\tilde{\bm{\phi}}_i(k-1)\bm{\mu}^{\rm T}_i(k-1)\}\{\Gamma^a_i(k)\}^{\rm T}\\
&\ \ \ \ \ \ -\Gamma^b_i(k){\rm E}\{\tilde{\bm{X}}_i(k-1)\bm{\mu}^{\rm T}_i(k-1)\}\{\Gamma^a_i(k)\}^{\rm T}\\
\end{aligned}
\end{eqnarray}
where $Q_i^a$ and $R_i$ are defined in (\ref{eq:n5}). By the definition of $\bm{\mu}_i(k-1)$, one has that
\begin{eqnarray}
\label{eq:100}
\begin{aligned}
&\ \ \ \ {\rm E}\{\bm{\mu}_i(k-1)\tilde{\bm{\phi}}^{\rm T}_i(k-1)\}\\
&={\rm E}\{\bm{\phi}_i(k)\tilde{\bm{\phi}}^{\rm T}_i(k-1)\}-{\rm E}\{\bm{\phi}_i(k-1)\tilde{\bm{\phi}}^{\rm T}_i(k-1)\}\\
\end{aligned}
\end{eqnarray}
Further, (\ref{eq:100}) can be rewritten as
\begin{eqnarray}
\label{eq:998}
\begin{aligned}
&\ \ \ \ {\rm E}\{\bm{\mu}_i(k-1)\tilde{\bm{\phi}}^{\rm T}_i(k-1)\}\\
&={\rm E}\{[\bm{\theta}_i(k)-\bm{\theta}_i(k-1)]\tilde{\bm{\phi}}^{\rm T}_i(k-1)\}\\
&\ \ \ -{\rm E}\{[\tilde{\bm{\phi}}_i(k-1)+\hat{\bm{\phi}}_i(k-1)]\tilde{\bm{\phi}}^{\rm T}_i(k-1)\}\\
&={\rm E}\{\bm{\theta}_i(k)\tilde{\bm{\phi}}^{\rm T}_i(k-1)\}-{\rm E}\{\bm{\theta}_i(k-1)\tilde{\bm{\phi}}^{\rm T}_i(k-1)\}\\
&\ \ \ -{\rm E}\{\hat{\bm{\phi}}_i(k-1)\tilde{\bm{\phi}}^{\rm T}_i(k-1)\}-P^\phi_{ii}(k-1)
\end{aligned}
\end{eqnarray}
on the basis of the definition of $\bm{\phi}_i(k)$ and $\tilde{\bm{\phi}}_i(k-1)$. Since $\hat{\bm{\phi}}_i(k-1)$ is designed in the linear minimum variance sense, one has that ${\rm E}\{\hat{\bm{\phi}}_i(k-1)\tilde{\bm{\phi}}^{\rm T}_i(k-1)\}=O_{p_i\times p_i}$. Meanwhile, when the condition (\ref{eq:666}) is valid, the term ${\rm E}\{\bm{\theta}_i(k-1)\tilde{\bm{\phi}}^{\rm T}_i(k-1)\}$ becomes
\begin{eqnarray}
{\rm E}\{\bm{\theta}_i(k-1)\bm{\theta}_i^{\rm T}(k-1)\}\{\Gamma^a_i(k-1)\}^{\rm T}
\end{eqnarray}
because $\tilde{\bm{\phi}}_i(k-1)$ can be calculated recursively by (\ref{eq:13}). Notice that ${\rm E}\{\bm{\theta}_i(k)\tilde{\bm{\phi}}^{\rm T}_i(k-1)\}=O_{p_i\times p_i}$ when the condition (\ref{eq:666}) holds. Then, it follows from the above analysis that
\begin{eqnarray}
\begin{aligned}
&\ \ \ \ {\rm E}\{\bm{\mu}_i(k-1)\tilde{\bm{\phi}}^{\rm T}_i(k-1)\}\\
&=-P^\phi_{ii}(k-1)-{\rm E}\{\bm{\theta}_i(k-1)\bm{\theta}_i^{\rm T}(k-1)\}\{\Gamma^a_i(k-1)\}^{\rm T}
\end{aligned}\;\;\;\;
\end{eqnarray}
At the same time, one has
\begin{eqnarray}
\label{eq:101}
\begin{aligned}
&\ \ \ \ {\rm E}\{\bm{\mu}_i(k-1)\tilde{\bm{X}}^{\rm T}_i(k-1)\}\\
&={\rm E}\{\bm{\phi}_i(k)\tilde{\bm{X}}^{\rm T}_i(k-1)\}-{\rm E}\{\bm{\phi}_i(k-1)\tilde{\bm{X}}^{\rm T}_i(k-1)\}\\
&={\rm E}\{\bm{\theta}_i(k)\tilde{\bm{X}}^{\rm T}_i(k-1)\}-{\rm E}\{\bm{\theta}_i(k-1)\tilde{\bm{X}}^{\rm T}_i(k-1)\}\\
&\ \ \ -U^{\rm T}_{ii}(k-1)-\Psi^{\rm T}_{ii}(k-1)
\end{aligned}
\end{eqnarray}
When (\ref{eq:666}) holds, it can also be derived that ${\rm E}\{\bm{\theta}_i(k)\tilde{\bm{X}}^{\rm T}_i(k-1)\}=O_{p_i\times (n+p_i)}$ and
\begin{eqnarray}
\begin{aligned}
&\ \ \ \ {\rm E}\{\bm{\theta}_i(k-1)\tilde{\bm{X}}^{\rm T}_i(k-1)\}\\
&={\rm E}\{\bm{\theta}_i(k-1)\bm{\theta}_i^{\rm T}(k-1)\}\{K^a_i(k-1)\Phi^a_i\}^{\rm T}
\end{aligned}
\end{eqnarray}
because $\tilde{\bm{X}}_i(k-1)$ can be calculated recursively by (\ref{eq:16}). Then, it can be obtained that
\begin{eqnarray}
\begin{aligned}
&\ \ \ \ {\rm E}\{\bm{\mu}_i(k-1)\tilde{\bm{X}}^{\rm T}_i(k-1)\}\\
&=-{\rm E}\{\bm{\theta}_i(k-1)\bm{\theta}_i^{\rm T}(k-1)\}\{K^a_i(k-1)\Phi^a_i\}^{\rm T}\\
&\ \ \  -U_{ii}^{\rm T}(k-1)-\Psi^{\rm T}_{ii}(k-1)
\end{aligned}
\end{eqnarray}
Furthermore, it is derived from (\ref{eq:666}) that
\begin{eqnarray}
\label{eq:995}
\begin{aligned}
&\ \ \ \ {\rm E}\{\bm{\mu}_i(k-1){\bm{\mu}}^{\rm T}_i(k-1)\}\\
&={\rm E}\{\bm{\theta}_i(k){\bm{\theta}}^{\rm T}_i(k)\}+4{\rm E}\{\bm{\theta}_i(k-1){\bm{\theta}}^{\rm T}_i(k-1)\}\\
&\ \ \ +{\rm E}\{\bm{\theta}_i(k-2){\bm{\theta}}^{\rm T}_i(k-2)\}
\end{aligned}
\end{eqnarray}
Note that $\bm{\theta}_i(k)$ is an unknown variable generated from the adversary, which means that it may not obey a probabilistic law. In this case, $\eta_i I_{p_i}$ is proposed to depict the term ${\rm E}\{\bm{\theta}_i(k)\bm{\theta}^{\rm T}_i(k)\}$. Substituting $\eta_i I_{p_i}$ for ${\rm E}\{\bm{\theta}_i(k){\bm{\theta}}^{\rm T}_i(k)\}$, ${\rm E}\{\bm{\theta}_i(k-1){\bm{\theta}}^{\rm T}_i(k-1)\}$ and ${\rm E}\{\bm{\theta}_i(k-2){\bm{\theta}}^{\rm T}_i(k-2)\}$, then the estimation error covariance matrix (\ref{eq:10}) is thus obtained.\\
\indent On the other hand, it follows from (\ref{eq:5}), (\ref{eq:16}) and the above analysis that
\begin{eqnarray}
\label{eq:17}
\begin{aligned}
&P^X_{ii}(k)=K_i(k)R_iK^{\rm T}_i(k)+K^a_i(k)[Q^a_i\\
&\ \ \  +A^a_i(k)P^X_{ii}(k-1)\{A^a_i(k)\}^{\rm T} \\
&\ \ \  -\Phi^a_iP^\phi_{ii}(k-1)\{\Phi^a_i\}^{\rm T}-A^a_i(k)U_{ii}(k-1)\{\Phi^a_i\}^{\rm T}\\
&\ \ \  -\Phi^a_iU^{\rm T}_{ii}(k-1)\{A^a_i(k)\}^{\rm T}+6\eta_i\Phi^a_i\{\Phi^a_i\}^{\rm T}\\
&\ \ \  -\eta_iA^a_i(k)K^a_i(k-1)\Phi^a_i\{\Phi^a_i\}^{\rm T}\\
&\ \ \  -\eta_i\Phi^a_i\Gamma^a_i(k-1)\{\Phi^a_i\}^{\rm T}-\eta_i\Phi^a_i\{\Phi^a_i\Gamma^a_i(k-1)\}^{\rm T}\\
&\ \ \  -\eta_i\Phi^a_i\{A^a_i(k)K^a_i(k-1)\Phi^a_i\}^{\rm T}]\{K^a_i(k)\}^{\rm T}\\
\end{aligned}
\end{eqnarray}
Hence, (\ref{eq:11}) is obtained from (\ref{eq:17}). Meanwhile, it is deduced from (\ref{eq:5}), (\ref{eq:99}) and (\ref{eq:16}) that
\begin{eqnarray}
\begin{aligned}
&{U_{ii}}(k) = K_i^a(k)A_i^a(k)[{U_{ii}}(k - 1){\{ \Gamma _i^a(k)\} ^{\rm{T}}}\\
&\;\;\;\;\;\; - {\eta _i}K_i^a(k - 1)\Phi _i^a{\{ \Phi _i^a\} ^{\rm{T}}}{\{ C_i^a(k)\} ^{\rm{T}}}\Gamma _i^{\rm{T}}(k)\\
&\;\;\;\;\;\; + P_{ii}^X(k - 1){\{ \Gamma _i^b(k)\} ^{\rm{T}}}] - K_i^a(k)\Phi _i^a{\{\Gamma _i^b(k)\Xi^2_i(k)\} ^{\rm{T}}}\\
&\;\;\;\;\;\; + K_i^a(k)\Phi _i^a\Xi^1_i(k){\{ \Phi _i^a\} ^{\rm{T}}}{\{ C_i^a(k)\} ^{\rm{T}}}\Gamma _i^{\rm{T}}(k)\\
&\;\;\;\;\;\; + K_i^a(k)Q_i^a{\{ C_i^a(k)\} ^{\rm{T}}}\Gamma _i^{\rm{T}}(k) - {K_i}(k){R_i}\Gamma _i^{\rm{T}}(k)\\
&\;\;\;\;\;\; + K_i^a(k)\Phi _i^a{\rm{E}}\{ {\mu _i}(k - 1)\hat \phi _i^{\rm{T}}(k - 1)\}
\end{aligned}
\end{eqnarray}
where $\Xi^1_i(k) \triangleq 6{\eta _i}{I_{{p_i}}} -P_{ii}^\phi (k - 1) - {\eta _i}\Gamma _i^a(k - 1) - {\eta _i}{\{ \Gamma _i^a(k - 1)\} ^{\rm{T}}}$, $\Xi^2_i(k)\triangleq U_{ii}(k-1)+\eta_i{K^a_i}(k-1){\Phi^a_i}$ and
\begin{eqnarray}
\begin{aligned}
&\ \ \ \ {\rm E}\{\bm{\mu}_i(k-1)\hat{\bm{\phi}}^{\rm T}_i(k-1)\}\\
&={\rm E}\{\bm{\theta}_i(k)\hat{\bm{\phi}}^{\rm T}_i(k-1)\}-{\rm E}\{\bm{\theta}_i(k-1)\hat{\bm{\phi}}^{\rm T}_i(k-1)\}\\
&\ \ \ -V_{ii}(k-1)\\
&=-{\rm E}\{\bm{\theta}_i(k-1)\bm{\theta}^{\rm T}_i(k-1)\}\{\Gamma_i(k-1)C^a_i(k-1)\Phi^a_i\}^{\rm T}\\
&\ \ \ -V_{ii}(k-1)
\end{aligned}\;\;\;
\end{eqnarray}
Taking place of ${\rm E}\{\bm{\theta}_i(k-1)\bm{\theta}^{\rm T}_i(k-1)\}$ by $\eta_iI_{p_i}$, (\ref{eq:12}) is derived. Finally, according to the definition and the above analysis, one can derive that
\begin{eqnarray}
\label{eq:102}
\begin{aligned}
&V_{ii}(k)=\Gamma_i(k)[R_i+C^a_i(k)Q_i^a\{C^a_i(k)\}^{\rm T}]\Gamma^{\rm T}_i(k)\\
&   +\Gamma^a_i(k)U^{\rm T}_{ii}(k-1)\{\Gamma^b_i(k)\}^{\rm T}+\Gamma^b_i(k)U_{ii}(k-1)\{\Gamma^a_i(k)\}^{\rm T}\\
&   +\Gamma^b_i(k)P^X_{ii}(k-1)\{\Gamma^b_i(k)\}^{\rm T}+V_{ii}(k-1)\{\Gamma^a_i(k)\}^{\rm T}\\
&   -\Gamma_i(k)C^a_i(k){\Phi^a_i}V_{ii}(k-1)-\Gamma_i(k)C^a_i(k){\Phi^a_i}\\
&   \times[P^\phi_{ii}(k-1)-6\eta_iI_{p_i}+\eta_i\Gamma^a_i(k-1)\\
&   +\eta_i\{\Gamma^a_i(k-1)\}^{\rm T}]\{\Phi^a_i\}^{\rm T}\{C^a_i(k)\}^{\rm T}\Gamma^{\rm T}_i(k)\\
&   -\eta_i\Gamma_i(k-1)C^a_i(k-1){\Phi^a_i}\{\Phi^a_i\}^{\rm T}\{C^a_i(k)\}^{\rm T}\Gamma^{\rm T}_i(k)\\
&   -\eta_i\Gamma^b_i(k)K^a_i(k-1){\Phi^a_i}\{\Phi^a_i\}^{\rm T}\{C^a_i(k)\}^{\rm T}\Gamma^{\rm T}_i(k)\\
&   -\eta_i\Gamma_i(k)C^a_i(k){\Phi^a_i}\{\Gamma^b_i(k)K^a_i(k-1)\Phi^a_i\}^{\rm T}\\
&   -\eta_i\Gamma_i(k)C^a_i(k){\Phi^a_i}\{\Phi^a_i\}^{\rm T}\{C^a_i(k-1)\}^{\rm T}\Gamma_i^{\rm T}(k-1)\\
\end{aligned}
\end{eqnarray}
which means that (\ref{eq:103}) holds. This completes the proof.

\textbf{Remark 5.} Under the condition (\ref{eq:666}) that information of attacks is unavailable,
the parameter $\eta_i$ is proposed to compensate the unknown term ${\rm E}\{\bm{\theta}_i(k)\bm{\theta}^{\rm T}_i(k)\}$. In this sense, $\eta_i$ is called as the compensation factor.
Generally, since the attack signal is unknown, the compensation factor can be used as an adjustable parameter to improve the estimation accuracy.


Based on Lemma 1, we shall obtain the following results.

\textbf{Theorem 1.} Given the compensation factor $\eta_i\ge0$. When the matrices $P^\phi_{ii}(k-1)$, $P^X_{ii}(k-1)$, $U_{ii}(k-1)$, $V_{ii}(k-1)$ are obtained from Lemma 1.
{The estimator gains $\Gamma_i(k)$ and $K_i(k)$ 
calculated by the following recursive form are optimal in the linear minimum variance sense:}
\begin{eqnarray}
\label{eq:7}
\begin{aligned}
&\Gamma_i(k)=-[P^\phi_{ii}(k-1)\{\Phi^a_i\}^{\rm T}+U_{ii}^{\rm T}(k-1)\{A^a_i(k)\}^{\rm T}\\
&\ \ \ \ \ \ \ \ +\eta_i{\Gamma_i^a}(k-1)\{\Phi^a_i\}^{\rm T}+\eta_i\{\Phi^a_i\Gamma_i^a(k-1)\}^{\rm T}\\
&\ \ \ \ \ \ \ \ +\eta_i\{A^a_i(k)K_i^a(k-1)\Phi^a_i\}^{\rm T}-6\eta_i\{\Phi^a_i\}^{\rm T}]\\
&\ \ \ \ \ \ \ \ \times\{C^a_i(k)\}^{\rm T}[C^a_i(k)\Xi_i(k)\{C^a_i(k)\}^{\rm T}+R_i]^{-1}
\end{aligned}\;\;\;\;\;\;\;\;
\end{eqnarray}
\begin{eqnarray}
\label{eq:8}
\begin{aligned}
&K_i(k)=\Xi_i(k)\{C^a_i(k)\}^{\rm T}[C^a_i(k)\Xi_i(k)\{C^a_i(k)\}^{\rm T}+R_i]^{-1}
\end{aligned}
\end{eqnarray}
where $\Xi_i(k)$ is defined by (\ref{eq:9}).

\textbf{Proof.} Taking the partial differentiation of $\mathrm{Tr}\{P^\phi_{ii}(k)\}$ with respect to $\Gamma_i(k)$ yields that
\begin{eqnarray}
\label{eq:15}
\begin{aligned}
&\ \ \ \ \partial \mathrm{Tr}\{P^\phi_{ii}(k)\}/\partial \Gamma_i(k)\\
&=2\Gamma_i(k)[C^a_i(k)\Xi_i(k)\{C^a_i(k)\}^{\rm T}+R_i]\\
&\ \   +2[P^\phi_{ii}(k-1)\{\Phi^a_i\}^{\rm T}+U_{ii}^{\rm T}(k-1)\{A^a_i(k)\}^{\rm T}\\
&\ \  +\eta_i{\Gamma_i^a}(k-1)\{\Phi^a_i\}^{\rm T}+\eta_i\{\Phi^a_i\Gamma_i^a(k-1)\}^{\rm T}\\
&\ \  +\eta_i\{A^a_i(k)K_i^a(k-1)\Phi^a_i\}^{\rm T}-6\eta_i\{\Phi^a_i\}^{\rm T}]\{C^a_i(k)\}^{\rm T}
\end{aligned}
\end{eqnarray}
where $P^\phi_{ii}(k)$ is given by (\ref{eq:10}) and $\Xi_i(k)$ is defined by (\ref{eq:9}). Let $\partial \mathrm{Tr}\{P^\phi_{ii}(k)\}/\partial \Gamma_i(k)=0$, the local estimator gain $\Gamma_i(k)$ can be computed by (\ref{eq:7}).\\
\indent
On the  other hand, taking the partial differentiation of $\mathrm{Tr}\{P^X_{ii}(k)\}$ with respect to $K_i(k)$ yields that
\begin{eqnarray}
\label{eq:18}
\begin{aligned}
&\ \ \ \ \partial \mathrm{Tr}\{P^X_{ii}(k)\}/\partial K_i(k)\\
&=2K_i(k)[C^a_i(k)\Xi_i(k)\{C^a_i(k)\}^{\rm T}+R_i]-2\Xi_i(k)\{C^a_i(k)\}^{\rm T}
\end{aligned}\;
\end{eqnarray}
where $P^X_{ii}(k)$ is given by (\ref{eq:11}). Let $\partial \mathrm{Tr}\{P^X_{ii}(k)\}/\partial K_i(k)=0$, the local estimator gain $K_i(k)$ is given by (\ref{eq:8}).\\
\indent
{To demonstrate that the estimator gain $K_i(k)$ derived by (\ref{eq:8}) makes the estimation error variance minimum, let $K_i^o(k)$ be the gain derived by (\ref{eq:8}) and $A_r$ be an arbitrary non-zero matrix with appropriate dimensions. Then, substituting $K_i^o(k)$ and $K_i^o(k)+A_r$ into (\ref{eq:11}) yields that
\begin{eqnarray}
\label{eq:n2}
\begin{aligned}
&\ \ \ \ P^X_{ii}(k)|_{K_i(k)=K_i^o(k)}\\
&=K_i^o(k)[R_i+C^a_i(k)\Xi_i(k)\{C^a_i(k)\}^{\rm T}]\{K^o_i(k)\}^{\rm T}\\
&\ +\Xi_i(k)-\Xi_i(k)\{K^o_i(k)C^a_i(k)\}^{\rm T}-K_i^o(k)C^a_i(k)\Xi_i(k)\\
\end{aligned}\;\;
\end{eqnarray}
\begin{eqnarray}
\label{eq:n1}
\begin{aligned}
&\ \ \ \ P^X_{ii}(k)|_{K_i(k)=K_i^o(k)+A_r}\\
&=\Xi_i(k)+[K_i^o(k)+A_r][R_i+C^a_i(k)\Xi_i(k)\{C^a_i(k)\}^{\rm T}]\\
&\ \times[K_i^o(k)+A_r]^{\rm T}-\Xi_i(k)\{[K_i^o(k)+A_r]C^a_i(k)\}^{\rm T}\\
&\ -[K_i^o(k)+A_r]C^a_i(k)\Xi_i(k)\\
\end{aligned}\;\;
\end{eqnarray}}\\
{From (\ref{eq:n2}) and (\ref{eq:n1}), the following equation can be obtained}
{\begin{eqnarray}
\begin{aligned}
&\ \ \ P^X_{ii}(k)|_{K_i(k)=K_i^o(k)+A_r}-P^X_{ii}(k)|_{K_i(k)=K_i^o(k)}\\
&=K_i^o(k)[R_i+C^a_i(k)\Xi_i(k)\{C^a_i(k)\}^{\rm T}]A_r^{\rm T}\\
&\ +A_r[R_i+C^a_i(k)\Xi_i(k)\{C^a_i(k)\}^{\rm T}]\{K_i^o(k)\}^{\rm T}\\
&\ +A_r[R_i+C^a_i(k)\Xi_i(k)\{C^a_i(k)\}^{\rm T}]A_r^{\rm T}\\
&\ -\Xi_i(k)\{A_rC^a_i(k)\}^{\rm T}-A_rC^a_i(k)\Xi_i(k)\\
\end{aligned}\;\;
\end{eqnarray}\\
Substituting $K_i^o(k)$ by (\ref{eq:8}) leads to that
\begin{eqnarray}
\begin{aligned}
&\ \ \ P^X_{ii}(k)|_{K_i(k)=K_i^o(k)+A_r}-P^X_{ii}(k)|_{K_i(k)=K_i^o(k)}\\
&=A_r[R_i+C^a_i(k)\Xi_i(k)\{C^a_i(k)\}^{\rm T}]A_r^{\rm T}
\end{aligned}\;\;
\end{eqnarray}\\
where
\begin{eqnarray}
\begin{aligned}
&\Xi_i(k)=\mathrm E\{[A^a_i(k)\tilde{X}_i(k-1)+\Phi^a_i\tilde{\bm{\phi}}_i(k-1)\\
&\ \ \ \ \ \ \ \ \ \ +\Phi^a_i\bm{\mu}_i(k-1)+W_i(k-1)]\\
&\ \ \ \ \ \ \ \ \ \ \times[A^a_i(k)\tilde{X}_i(k-1)+\Phi^a_i\tilde{\bm{\phi}}_i(k-1)\\
&\ \ \ \ \ \ \ \ \ \ +\Phi^a_i\bm{\mu}_i(k-1)+W_i(k-1)]^\mathrm T\}\ge 0
\end{aligned}
\end{eqnarray}
Hence, $P^X_{ii}(k)|_{K_i(k)=K_i^o(k)+A_r}>P^X_{ii}(k)|_{K_i(k)=K_i^o(k)}$ for any arbitrary non-zero matrix $A_r$. Thus, $K_i^o(k)$ is the only extreme point of $\mathrm {Tr}\{P_{ii}^X(k)\}$ with respect to $K_i(k)$, and $K_i(k)$ given by (\ref{eq:8}) can minimize $\mathrm {Tr}\{P_{ii}^X(k)\}$. Similarly, let $\Gamma_i^o(k)$ be the estimator gain derived by (\ref{eq:7}) and $B_r$ be an arbitrary non-zero matrix with appropriate dimensions. By substituting them into (\ref{eq:10}),}
{it is found that $P^\phi_{ii}(k)|_{\Gamma_i(k)=\Gamma_i^o(k)+B_r}>P^\phi_{ii}(k)|_{\Gamma_i(k)=\Gamma_i^o(k)}$ for any arbitrary non-zero matrix $B_r$. Thus, $\Gamma_i^o(k)$ is the only extreme point of $\mathrm {Tr}\{P_{ii}^\phi(k)\}$ with respect to $\Gamma_i(k)$, and $\Gamma_i(k)$ given by (\ref{eq:7}) can minimize $\mathrm {Tr}\{P_{ii}^\phi(k)\}$. This means that the designed estimator gains are optimal in the linear minimum variance sense.} This completes the proof.

\textbf{Remark 6.} Though the estimator structure (\ref{eq:99}) is similar with Eq. (5k) and Eq. (5l) in \cite{c95}, the design of estimator gains $K_i(k)$ and $\Gamma_i(k)$ in this paper are different from that of \cite{c95}. Specifically, the adaptive Kalman filter in \cite{c95} was designed based on the condition that the unknown input was constant, thus the developed method in (\ref{eq:99}) is suitable for the case that the unknown input is time-invariant or it varies extremely slowly. In contrast, the proposed method in Theorem 1 takes the variability of attack signals into consideration, and the proposed compensation factor can enable the designed secure estimator to perform well under the condition that the unknown input is time varying. At the same time, the advantages of the proposed method has been demonstrated by comparing with the method of \cite{c95} in Simulations.

Next, the estimation error cross-covariance matrix between two local estimators will be determined by Theorem 2.

\indent \textbf{Theorem 2.} Under the initial values $P^\phi_{ij}(0)$, $P^X_{ij}(0)$, $U_{ij}(0)$, $Y_{ij}(0)$ and $V_{ij}(0)\ (i\ne j)$. When each local estimator gains $K_i(k)$, $\Gamma_i(k)$ are given in Theorem 1, the estimation error cross-covariance matrices can be calculated by the following recursive form:
\begin{eqnarray}
\label{eq:23}
\begin{aligned}
&P^\phi_{ij}(k)=\Gamma_i(k)C^a_i(k)\Phi^a_i\Xi^1_{ij}(k)-\Xi^1_{ij}(k)\{\Gamma^a_j(k)\}^{\rm T}\\
&\ \ \ \ \ \ \ \     +U^{\rm T}_{ji}(k-1)\{\Gamma^b_j(k)\}^{\rm T}+\Gamma^b_i(k)U_{ij}(k-1)\\
&\ \ \ \ \ \ \ \    +\Gamma_i(k)C^a_i(k)\Xi_{ij}(k)\{\Gamma_j(k)C^a_j(k)\}^{\rm T}\\
\end{aligned}\;\;\;\;
\end{eqnarray}
\begin{eqnarray}
\label{eq:24}
\begin{aligned}
&P^X_{ij}(k)=K^a_i(k)\Xi_{ij}(k)\{K^a_j(k)\}^{\rm T}
\end{aligned}
\end{eqnarray}
\begin{eqnarray}
\label{eq:25}
\begin{aligned}
&U_{ij}(k)=K^a_i(k)[A^a_i(k)U_{ij}(k-1)-\Phi^a_iV_{ij}(k-1)]\\
&\ \ \ \ \ \ \ \   +K^a_i(k)\Xi_{ij}(k)\{\Gamma_j(k)C^a_j(k)\}^{\rm T}\\
\end{aligned}\;\;\;\;
\end{eqnarray}
\begin{eqnarray}
\label{eq:106}
\begin{aligned}
&Y_{ij}(k)=-U^{\rm T}_{ji}(k-1)\{\Gamma^b_j(k)\}^{\rm T}-\Gamma^b_i(k)U_{ij}(k-1)\\
&\ \ \ \     -\Xi^1_{ij}(k)\{\Gamma_j(k)C^a_j(k)\Phi^a_j\}^{\rm T}-\Gamma^a_i(k)V_{ij}(k-1)\\
&\ \ \ \     -\Gamma_i(k)C^a_{i}(k)\Xi_{ij}(k)\{\Gamma_j(k)C^a_j(k)\}^{\rm T}
\end{aligned}\;\;\;\;
\end{eqnarray}
\begin{eqnarray}
\label{eq:105}
\begin{aligned}
&V_{ij}(k)=\Gamma_i(k)C^a_i(k)\Xi_{ij}(k)\{\Gamma_j(k)C^a_j(k)\}^{\rm T}\\
&\ \ \  +V_{ij}(k-1)\{\Gamma^a_j(k)\}^{\rm T}-\Gamma_i(k)C^a_i(k){\Phi^a_i}V_{ij}(k-1)\\
&\ \ \  +U_{ji}^{\rm T}(k-1)\{\Gamma^b_j(k)\}^{\rm T}+\Gamma^b_i(k)U_{ij}(k-1)\\
\end{aligned}\;
\end{eqnarray}
where
\begin{eqnarray}
\begin{cases}
\begin{aligned}
&\Xi_{ij}(k)\triangleq A^a_i(k)P^X_{ij}(k-1)\{A^a_j(k)\}^{\rm T}\\
&\ \ \ \ \ \ \ \ \ \ \ \ -A^a_i(k)U_{ij}(k-1)\{\Phi^a_j\}^{\rm T}\\
&\ \ \ \ \ \ \ \ \ \ \ \ -\Phi^a_iU_{ji}^{\rm T}(k-1)\{A^a_j(k)\}^{\rm T}\\
&\ \ \ \ \ \ \ \ \ \ \ \ -\Phi^a_i\Xi^1_{ij}(k)\{\Phi^a_j\}^{\rm T}+Q^a_{ij}\\
&\Xi^1_{ij}(k)\triangleq P^\phi_{ij}(k-1)+Y_{ij}(k-1)+Y^{\rm T}_{ji}(k-1)
\end{aligned}
\end{cases}
\end{eqnarray}
and $Q_{ij}^a$ is defined in (\ref{eq:n5}).

\textbf{Proof.}
According to (\ref{eq:5}) and (\ref{eq:13}), the estimation error cross-covariance matrix $P^\phi_{ij}(k)$ is given by
\vspace{-5pt}
\begin{eqnarray}
\begin{aligned}
&P^\phi_{ij}(k)=\Gamma_i(k)C^a_i(k)Q^a_{ij}\{C^a_j(k)\}^{\rm T}\Gamma_j^{\rm T}(k)\\
&\ \ \ \ \  +\Gamma^a_i(k)P^\phi_{ij}(k-1)\{\Gamma^a_j(k)\}^{\rm T}\\
&\ \ \ \ \  +\Gamma^b_i(k)P^X_{ij}(k-1)\{\Gamma^b_j(k)\}^{\rm T}\\
&\ \ \ \ \  -\Gamma^a_i(k)\Psi^{\rm T}_{ji}(k-1)\{\Gamma^b_j(k)\}^{\rm T}\\
&\ \ \ \ \  -\Gamma^b_i(k)\Psi_{ij}(k-1)\{\Gamma^a_j(k)\}^{\rm T}\\
&\ \ \ \ \  +\Gamma^a_i(k){\rm E}\{\bm{\mu}_i(k-1)\bm{\mu}^{\rm T}_j(k-1)\}\{\Gamma^a_j(k)\}^{\rm T}\\
&\ \ \ \ \  +\Gamma^a_i(k){\rm E}\{\bm{\mu}_i(k-1)\tilde{\bm{\phi}}^{\rm T}_j(k-1)\}\{\Gamma^a_j(k)\}^{\rm T}\\
&\ \ \ \ \  -\Gamma^a_i(k){\rm E}\{\bm{\mu}_i(k-1)\tilde{\bm{X}}^{\rm T}_j(k-1)\}\{\Gamma^b_j(k)\}^{\rm T}\\
&\ \ \ \ \  +\Gamma^a_i(k){\rm E}\{\tilde{\bm{\phi}}_i(k-1)\bm{\mu}^{\rm T}_j(k-1)\}\{\Gamma^a_j(k)\}^{\rm T}\\
&\ \ \ \ \  -\Gamma^b_i(k){\rm E}\{\tilde{\bm{X}}_i(k-1)\bm{\mu}^{\rm T}_j(k-1)\}\{\Gamma^a_j(k)\}^{\rm T}\\
\end{aligned}
\end{eqnarray}
where $Q_{ij}^a$ is defined in (\ref{eq:n5}). According to (\ref{eq:998}), one has that
\begin{eqnarray}
\begin{aligned}
&\ \ \ \ {\rm E}\{\bm{\mu}_i(k-1)\tilde{\bm{\phi}}^{\rm T}_j(k-1)\}\\
&={\rm E}\{\bm{\theta}_i(k)\tilde{\bm{\phi}}^{\rm T}_j(k-1)\}-{\rm E}\{\bm{\theta}_i(k-1)\tilde{\bm{\phi}}^{\rm T}_j(k-1)\}\\
&\ \ \ -Y^\mathrm T_{ji}(k-1)-P^\phi_{ij}(k-1)
\end{aligned}
\end{eqnarray}
When the condition (\ref{eq:666}) holds, one has ${\rm E}\{\bm{\theta}_i(k)\tilde{\bm{\phi}}^{\rm T}_j(k-1)\}=O_{p_i\times p_j}$, and ${\rm E}\{\bm{\theta}_i(k-1)\tilde{\bm{\phi}}^{\rm T}_j(k-1)\}$ becomes
\begin{eqnarray}
P^\theta_{ij}(k-1)\{\Gamma^a_j(k-1)\}^{\rm T}=O_{p_i\times p_j}
\end{eqnarray}
where $P^\theta_{ij}(k)$ is defined in (\ref{eq:5}). Then, it follows from the above analysis that
\begin{eqnarray}
\begin{aligned}
{\rm E}\{\bm{\mu}_i(k-1)\tilde{\bm{\phi}}^{\rm T}_j(k-1)\}=-P^\phi_{ij}(k-1)-Y^\mathrm T_{ji}(k-1)
\end{aligned}\;\;\;
\end{eqnarray}
At the same time, it is obtained from (\ref{eq:101}) that
\begin{eqnarray}
\label{eq:a101}
\begin{aligned}
&\ \ \ \ {\rm E}\{\bm{\mu}_i(k-1)\tilde{\bm{X}}^{\rm T}_j(k-1)\}\\
&={\rm E}\{\bm{\theta}_i(k)\tilde{\bm{X}}^{\rm T}_j(k-1)\}-{\rm E}\{\bm{\theta}_i(k-1)\tilde{\bm{X}}^{\rm T}_j(k-1)\}\\
&\ \ \ -U^{\rm T}_{ji}(k-1)-\Psi^{\rm T}_{ji}(k-1)
\end{aligned}
\end{eqnarray}
When (\ref{eq:666}) holds, it can also be derived that ${\rm E}\{\bm{\theta}_i(k)\tilde{\bm{X}}^{\rm T}_j(k-1)\}=O_{p_i\times (n+p_j)}$ and
\begin{eqnarray}
\label{eq:b101}
\begin{aligned}
&\ \ \ \ {\rm E}\{\bm{\theta}_i(k-1)\tilde{\bm{X}}^{\rm T}_j(k-1)\}\\
&=P^\theta_{ij}(k-1)\{K^a_j(k-1)\Phi^a_j\}^{\rm T}=O_{p_i\times (n+p_j)}
\end{aligned}
\end{eqnarray}
Then, (\ref{eq:a101}) can be rewritten as
\begin{eqnarray}
\begin{aligned}
{\rm E}\{\bm{\mu}_i(k-1)\tilde{\bm{X}}^{\rm T}_j(k-1)\}=-U_{ji}^{\rm T}(k-1)-\Psi^{\rm T}_{ji}(k-1)
\end{aligned}\;\;
\end{eqnarray}
Furthermore, by (\ref{eq:995}) and the first equation in (\ref{eq:666}) one has
\begin{eqnarray}
\begin{aligned}
&\ \ \ \ {\rm E}\{\bm{\mu}_i(k-1){\bm{\mu}}^{\rm T}_j(k-1)\}\\
&=P^\theta_{ij}(k)+4P^\theta_{ij}(k-1)+P^\theta_{ij}(k-2)=O_{p_i\times p_j}
\end{aligned}
\end{eqnarray}
Then, the estimation error cross-covariance matrix (\ref{eq:23}) is thus derived. On the other hand, it follows from (\ref{eq:5}), (\ref{eq:16}) and the above analysis that
\begin{eqnarray}
\begin{aligned}
&P^X_{ij}(k)=K^a_i(k)[A^a_i(k)P^X_{ij}(k-1)\{A^a_j(k)\}^{\rm T}\\
&\      -A^a_i(k)U_{ij}(k-1)\{\Phi^a_j\}^{\rm T}-{\Phi^a_i}Y_{ij}(k-1)\{\Phi^a_j\}^{\rm T}\\
&\      -\Phi^a_iU_{ji}^{\rm T}(k-1)\{A^a_j(k)\}^{\rm T}-{\Phi^a_i}Y_{ji}^{\rm T}(k-1)\{\Phi^a_j\}^{\rm T}\\
&\      -\Phi^a_iP^\phi_{ij}(k-1)\{\Phi^a_j\}^{\rm T}+Q^a_{ij}]\{K^a_j(k)\}^{\rm T}
\end{aligned}
\end{eqnarray}
Hence, (\ref{eq:24}) is obtained. Meanwhile, it is deduced from (\ref{eq:99}), (\ref{eq:5}) and (\ref{eq:16}) that
\begin{eqnarray}
\label{eq:19}
\begin{aligned}
&U_{ij}(k)=K^a_i(k)\Phi^a_i[-Y^{\rm T}_{ji}(k-1)-P^\phi_{ij}(k-1)]\\
&\ \ \ \ \ \ \ \ \ \ \   \times\{\Phi^a_j\}^{\rm T}\{C^a_j(k)\}^{\rm T}\Gamma_j^{\rm T}(k)\\
&\ \ \ \ \ \ \ \ \ \ \   +K^a_i(k)A^a_i(k)[U_{ij}(k-1)\{\Gamma^a_j(k)\}^{\rm T}\\
&\ \ \ \ \ \ \ \ \ \ \   +P^X_{ij}(k-1)\{\Gamma^b_j(k)\}^{\rm T}]+K^a_i(k)\Phi^a_i\\
&\ \ \ \ \ \ \ \ \ \ \   \times[Y_{ij}(k-1)\{\Gamma^a_j(k)\}^{\rm T}\\
&\ \ \ \ \ \ \ \ \ \ \   -U^{\rm T}_{ji}(k-1)\{\Gamma^b_j(k)\}^{\rm T}]\\
&\ \ \ \ \ \ \ \ \ \ \   +K^a_i(k)Q_{ij}^a\{C^a_j(k)\}^{\rm T}\Gamma_j^{\rm T}(k)\\
&\ \ \ \ \ \ \ \ \ \ \   +K^a_i(k)\Phi^a_i{\rm E}\{\bm{\mu}_i(k-1)\hat{\bm{\phi}}^{\rm T}_j(k-1)\}\\
\end{aligned}\;\;\;
\end{eqnarray}
where
\begin{eqnarray}
\begin{aligned}
&\ \ \ \ {\rm E}\{\bm{\mu}_i(k-1)\hat{\bm{\phi}}^{\rm T}_j(k-1)\}\\
&={\rm E}\{\bm{\theta}_i(k)\hat{\bm{\phi}}^{\rm T}_j(k-1)\}-{\rm E}\{\bm{\theta}_i(k-1)\hat{\bm{\phi}}^{\rm T}_j(k-1)\}\\
&\ \ \ -Y_{ij}(k-1)-V_{ij}(k-1)\\
&=-Y_{ij}(k-1)-V_{ij}(k-1)
\end{aligned}
\end{eqnarray}
Thus, (\ref{eq:25}) is obtained from (\ref{eq:19}). Finally, according to the definition and the above analysis, one can derive that
\begin{eqnarray}
\begin{aligned}
&Y_{ij}(k)=\Gamma^a_i(k)[-Y^{\rm T}_{ji}(k-1)-Y_{ij}(k-1)\\
&\ \ \ \ \ \ \ \ \ \  -P^\phi_{ij}(k-1)]\{\Phi^a_j\}^{\rm T}\{C^a_j(k)\}^{\rm T}\Gamma_j^{\rm T}(k)\\
&\ \ \ \ \ \ \ \ \ \  -\Gamma^a_i(k)V_{ij}(k-1)\\
&\ \ \ \ \ \ \ \ \ \  -\Gamma^a_i(k)U^{\rm T}_{ji}(k-1)\{\Gamma^b_j(k)\}^{\rm T}\\
&\ \ \ \ \ \ \ \ \ \  -\Gamma^b_i(k)U_{ij}(k-1)\{\Gamma^a_j(k)\}^{\rm T}\\
&\ \ \ \ \ \ \ \ \ \  -\Gamma^b_i(k)P^X_{ij}(k-1)\{\Gamma^b_j(k)\}^{\rm T}\\
&\ \ \ \ \ \ \ \ \ \  -\Gamma_i(k)C^a_{i}(k)Q_{ij}^a\{C^a_j(k)\}^{\rm T}\Gamma^{\rm T}_j(k)
\end{aligned}\;\;\;\;\;
\end{eqnarray}
\begin{eqnarray}
\begin{aligned}
&V_{ij}(k)=-\Gamma_i(k)C^a_i(k){\Phi^a_i}[Y_{ij}(k-1)+Y^{\rm T}_{ji}(k-1)\\
&\ \ \ \ \ \ \ \ \ \ \ +P^\phi_{ij}(k-1)]\{\Phi^a_j\}^{\rm T}\{C^a_j(k)\}^{\rm T}\Gamma^{\rm T}_j(k)\\
&\ \ \ \ \ \ \ \ \ \ \ +V_{ij}(k-1)\{\Gamma^a_j(k)\}^{\rm T}\\
&\ \ \ \ \ \ \ \ \ \ \ -\Gamma_i(k)C^a_i(k){\Phi^a_i}V_{ij}(k-1)\\
&\ \ \ \ \ \ \ \ \ \ \ +\Gamma^a_i(k)U_{ji}^{\rm T}(k-1)\{\Gamma^b_j(k)\}^{\rm T}\\
&\ \ \ \ \ \ \ \ \ \ \ +\Gamma^b_i(k)U_{ij}(k-1)\{\Gamma^a_j(k)\}^{\rm T}\\
&\ \ \ \ \ \ \ \ \ \ \ +\Gamma^b_i(k)P^X_{ij}(k-1)\{\Gamma^b_j(k)\}^{\rm T}\\
&\ \ \ \ \ \ \ \ \ \ \ +\Gamma_i(k)C^a_i(k)Q_{ij}^a\{C^a_j(k)\}^{\rm T}\Gamma^{\rm T}_j(k)\\
\end{aligned}
\end{eqnarray}
Then, (\ref{eq:106}) and (\ref{eq:105}) are thus obtained. This completes the proof.

Based on Theorems 1 and 2, the computation procedures for the fusion estimate $\hat{\bm{x}}_0(k)$ of the state $\bm{x}(k)$ under Case I are shown by Algorithm 1.
\begin{algorithm}[h]
    \caption{Secure Fusion Estimation under Gaussian Noises}
   {\small{
    \begin{algorithmic}[1] 
        \STATE Set the compensation factors $\eta_i\;(i=1,2,\ldots,r)$.
        \FOR{$i := 1$ \TO $r$}{
            \STATE Calculate $K_i(k)$ and $\Gamma_i(k)$ by (\ref{eq:7}) and (\ref{eq:8});
             \STATE Calculate $\hat{\bm{X}}_i(k)$ and $\hat{\bm{\phi}}_i(k)$ by (\ref{eq:99}).
        }
        \ENDFOR
        \STATE Calculate $G(k)$ by (\ref{eq:21});
        \STATE Calculate $\hat{\bm{x}}_0(k)$ by (\ref{eq:6});
        \STATE Return to step 2 and implement steps 2-7 for obtaining $\hat{\bm{x}}_0(k+1)$.
    \end{algorithmic}}}
\end{algorithm}

\section{Simulation Examples}
Consider a power grid with IEEE 4-bus distribution line that adopts the model of interconnected distributed energy generators (DEGs). In this example, four DEGs are modeled as voltage sources whose input voltages are denoted as $\bm{v}_p\triangleq[v_{p1};v_{p2};v_{p3};v_{p4}]$, where $v_{pi}$ is the $i$th DEG input voltage. At the same time, the four DEGs are connected to the main power networks at the corresponding point of common coupling (PCC) whose voltages are denoted as $\bm{v}_s\triangleq[v_{1};v_{2};v_{3};v_{4}]$, where $v_{i}$ is the $i$th PCC voltages. To maintain the proper operation of DEGs, these PCC voltages need to be kept at their reference values, while a coupling inductor exists between each DEG and the rest of the electricity networks. Then, the nodal voltage equation can be converted into the following linear state-space dynamical model \cite{c994}:
\begin{equation}
\label{eq:38}
\dot{\bm{x}}(t)=A_c\bm{x}(t)+B_c\bm{u}(t)
\end{equation}
where $\bm{x}(t)\triangleq\bm{v}_s-\bm{v}_{\rm ref}$ is the PCC state voltage deviation, $\bm{v}_{\rm ref}$ is the PCC reference voltage, $\bm{u}(t)\triangleq\bm{v}_p-\bm{v}_{\rm pref}$ is the DEG control input deviation, $\bm{v}_{\rm pref}$ is the reference control effort. Here, the system matrices $A_c$ and $B_c$ are taken as \cite{c98}:
\begin{equation}
\label{eq:39}
A_c=
\left[ {\begin{array}{*{20}{c}}
175.9 & 176.8 & 511 & 1036
  \\ -350 & 0 & 0 & 0
  \\ -544.2 & -474.8 & -408.8 & -828.8
  \\ -119.7 & -554.6 & -968.8 & -1077.5
\end{array}} \right]\;\;\;\;
\end{equation}
\begin{equation}
\label{eq:40}
B_c=
\left[ {\begin{array}{*{20}{c}}
0.8 & 334.2 & 525.1 & -103.6
  \\ -350 & 0 & 0 & 0
  \\ -69.3 & -66.1 & -420.1 & -828.8
  \\ -434.9 & -414.2 & -108.7 & -1077.5
\end{array}} \right]\;\;\;\;
\end{equation}
Notice that the system (\ref{eq:38}) is unstable when there is no feedback control. Under this situation, the controller $\bm{u}(t)\triangleq K_c\bm{x}(t)$ is designed such that the system can be stable, i.e., all eigenvalues of $A_s \triangleq A_c + B_cK_c$ are negative. In this case, the controller gain $K_c$ is chosen as
\begin{equation}
\label{eq:41}
K_c=
\left[ {\begin{array}{*{20}{c}}
-1.0057 & 0 & 0 & 0
  \\ 1.2883 & -0.2003 & -1.4687 & -1.4687
  \\ -1.1696 & -0.2936 & -0.1024 & -1.1021
  \\ -0.0824 & -0.4081 & -0.3242 & -0.3242
\end{array}} \right]\;
\end{equation}
Then, the system (\ref{eq:38}) can be rewritten as
\begin{equation}
\label{eq:42}
\dot{\bm{x}}(t)=A_s\bm{x}(t)
\end{equation}
To monitor the work status of the power grid, five sensors are deployed to collect measurement information. By setting the sampling period $T= 5s$, (\ref{eq:42}) can be transformed to the same form of (\ref{eq:1}), where
\begin{equation}
\label{eq:43}
A=
\left[ {\begin{array}{*{20}{c}}
-0.837 & 0.5427 & 0 & 0
  \\ -0.5427 & -0.837 & 0 & 0
  \\ 0 & 0 & 0.9851 & 0
  \\ 0 & 0 & 0 & 0.9556
\end{array}} \right]
\end{equation}
and the covariance of the noise $\bm{w}(k)$ is taken as $Q = {\rm diag}\{0.1,0.2,0.3,0.2\}$. Then, the measurement matrices are taken as
\begin{equation*}
\begin{aligned}
&C^o_1=[ \begin{array}{*{20}{c}}
1&0&0&0
\end{array}],C^o_2=[ \begin{array}{*{20}{c}}
0&0&1&0
\end{array}],C^o_3=[ \begin{array}{*{20}{c}}
1&0&0&1
\end{array}]\\&C^o_4=[ \begin{array}{*{20}{c}}
0&0&1&1
\end{array}],C^o_5=[ \begin{array}{*{20}{c}}
0&1&1&0
\end{array}]
\end{aligned}
\end{equation*}
and the covariance of the measurement noises are taken as $R_1^o=R_2^o=R_3^o=R_4^o=R_5^o=0.1$. In this example, sensor 1 and sensor 2 are chosen as the weak-defense sensors while the others are strong-defense sensors. Then, the weak-defense sensors are combined with strong-defense sensors, and the augmented systems are constructed based on sensor 1 and sensor 2, which yields that

\vspace{-3pt}
\begin{equation}
\begin{cases}
{\bm{X}}_i(k)=A^a_i{\bm{X}}_i(k-1)+\Phi^a_i{\bm{\phi}}_i(k)
\\\ \ \ \ \ \ \ \ \ \ \ \ +{\bm{W}}_i(k-1)
\\{\bm{y}}_i(k)=C_i^a{\bm{X}}_i(k)+{\bm{v}}_i(k)
\end{cases}
\end{equation}
where
\begin{equation*}
A^a_1=A^a_2=
\left[ {\begin{array}{*{20}{c}}
-0.837 & 0.5427 & 0 & 0&0
  \\ -0.5427 & -0.837 & 0 & 0&0
  \\ 0 & 0 & 0.9851 & 0  &0
  \\ 0 & 0 & 0 & 0.9556&0
  \\ 0& 0& 0& 0& 1
\end{array}} \right]
\end{equation*}
and
\begin{equation*}
\begin{aligned}
&C^a_1=\left[ \begin{array}{*{20}{c}}
1&0&0&0&1\\
1&0&0&1&0\\
0&0&1&1&0
\end{array}\right],C^a_2=\left[ \begin{array}{*{20}{c}}
0&0&1&0&1\\
1&0&0&1&0\\
0&1&1&0&0
\end{array}\right]
\end{aligned}
\end{equation*}
In the simulation, the attack signal $\bm{\theta}_1(k)$ is the Gaussian white noise with covariance $5$ while the attack signal $\bm{\theta}_2(k)$ is taken as
\begin{equation*}
\begin{aligned}
&\bm{\theta}_2(k)=\left\{ \begin{array}{l}
0,\ \ 0\le k\le 49\\3,\ \ 50\le k<51\\0,\ \ 51\le k\le 100
\end{array}\right.
\end{aligned}
\end{equation*}
\begin{figure}
\centering
\includegraphics[height=5.0cm, width=8.5cm]{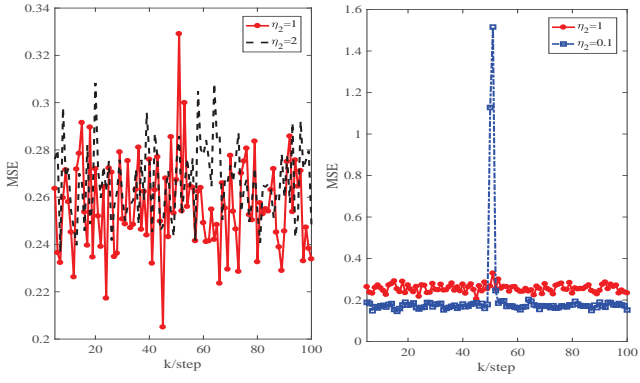}
\caption{The performance comparison of attack estimators for sensor 2 with different compensation factors $\eta_2$}
\label{figurelabel9}
\end{figure}
\begin{figure}
\centering
\includegraphics[height=5.0cm, width=8.5cm]{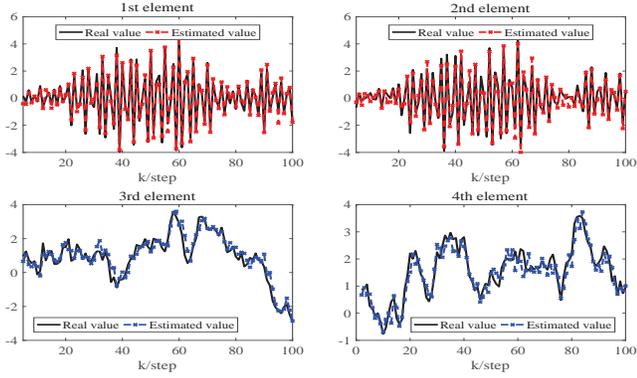}
\caption{The system state and its fusion estimate obtained by Algorithm 1. ($\eta_1=\eta_2=1$)}
\label{figurelabel12}
\end{figure}
\begin{figure}
\centering
\includegraphics[height=5.0cm, width=8.5cm]{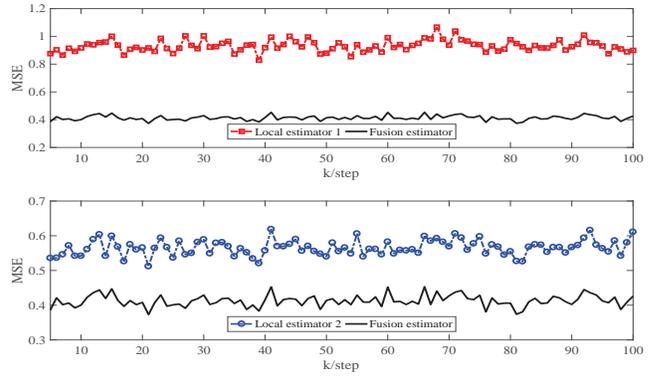}
\caption{The performance comparison of the local estimators and the fusion estimator given by Algorithm 1}
\label{figurelabel3}
\end{figure}
\begin{figure}
\centering
\includegraphics[height=5.0cm, width=8.5cm]{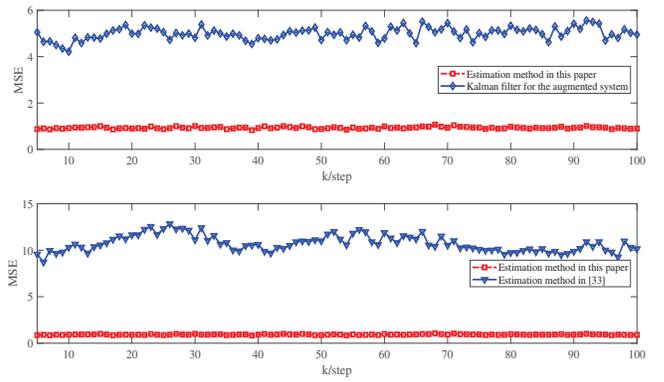}
\caption{The performance comparison of local estimators obtained by different methods for sensor 1}
\label{figurelabel10}
\end{figure}

By implementing Algorithm 1, Fig. \ref{figurelabel9} shows mean square errors (MSEs) of the attack estimator calculated by the Monte Carlo method with an average of 500 runs. From this figure, it is seen that the estimator has different performance as the compensation factor varies as stated in Remark 5. Thus, this urges us to design the selection criteria for the time varying compensation factor. Meanwhile, the real value of system state and its fusion estimate are plotted in Fig. \ref{figurelabel12}. It is seen from Fig. \ref{figurelabel12} that the fusion estimator given by Algorithm 1 can estimate the system state well. To compare the performance of the local estimators and the fusion estimator given by Algorithm 1, when choosing $\eta_1=\eta_2=1$, Fig. \ref{figurelabel3} shows the MSEs of state estimators calculated by the Monte Carlo method with an average of 500 runs. It is seen from Fig. \ref{figurelabel3} that the fusion estimator performs well for estimating the state, and the fusion estimator has less MSE than each local estimator. This accords with the expected performance of the fusion system.

To demonstrate the advantages of the proposed estimation algorithm, it is compared with the augmented Kalman filtering method in Remark 3 and the adaptive Kalman filtering method in \cite{c95}. Then, Fig. \ref{figurelabel10} shows the MSEs of different estimators calculated by the Monte Carlo method with an average of 500 runs for sensor 1. It is seen from Fig. \ref{figurelabel10}(a) that the estimation precision of the local estimator given by Algorithm 1 is higher than the augmented Kalman filter (see (\ref{eq:98})), which means that the proposed local estimator has better performance than the augmented Kalman filter under sensor attacks. This is because there is no statistical information of attacks for designing the Kalman filter gains. At the same time, Fig. \ref{figurelabel10}(b) shows the estimation performance of Algorithm 1 and the adaptive Kalman filter in \cite{c95}, and it is obvious that the designed local estimator in this paper has less MSE than the method in \cite{c95}. This verifies the result in Remark 6, i.e., when the unknown input is time-varying, the proposed local estimation method works well, but the performance of adaptive Kalman filtering method in \cite{c95} becomes worse.

\section{CONCLUSIONS}
This paper studied the secure state fusion estimation problem in CPSs, where sensor measurements may be tampered by FDI attacks. Considering that some sensors may not be attacked, the system was reconstructed by modelling the attack signals as elements of the state vector, while the difference of the attacks between the current moment and the previous moment became an unknown input. Then, the secure state estimation problem was formulated into the joint estimation problem of the augmented state and the unknown input. In this case, optimal local estimators and distributed fusion criteria were designed respectively. Finally, illustrative examples were used to testify the effectiveness of the proposed methods.  

\vfill
\end{document}